\documentclass[english,prd,superscriptaddress,nofootinbib,preprintnumbers,showpacs,preprint]{revtex4}
\usepackage[latin1]{inputenc}
\usepackage{graphicx}
\usepackage{amsmath,amsthm,amssymb}
\usepackage{bm}

\usepackage{amsfonts}
\usepackage{dcolumn}

\usepackage{color}

\newcommand{\N}{{\cal N}}

\newcommand{\beqa}{\begin{eqnarray}}
\newcommand{\eeqa}{\end{eqnarray}}

\newcommand{\siml}{\lesssim}

\usepackage{babel}
\begin{document}

\begin{flushright}
RUP-17-5

\end{flushright}

\title{Spin Distribution of Primordial Black Holes
}

\author{Takeshi Chiba}
\affiliation{Department of Physics, College of Humanities and Sciences, 
Nihon University, Tokyo 156-8550, Japan}
\author{Shuichiro Yokoyama}
\affiliation{Department of Physics,  Rikkyo University, 
Tokyo 171-8501, Japan}

\begin{abstract}
We estimate the spin distribution of primordial black holes 
based on the recent study of the critical phenomena in the gravitational collapse of a rotating radiation fluid. 
We find that primordial black holes are mostly  slowly rotating.
\end{abstract}

\date{\today}

\pacs{98.80.Cq, 98.80.Es}

\maketitle

\section{Introduction}

LIGO-VIRGO Collaboration has finally detected gravitational waves 
and also revealed the existence of  binary black holes (BHs) in our universe~\cite{Abbott:2016blz}.
In particular, the source of the first gravitational event, named GW150914, 
 was reported to be a merger of a binary system of BHs with mass 
$30 {\rm M}_\odot$, which is larger than that expected for a standard stellar BH. 
 The origins of such massive 
BH binaries are proposed,  ranging from an isolated 
stellar binary system to dense stellar clusters \cite{ligo2}, or even to 
primordial black holes (PBHs)~\cite{Clesse:2016vqa, 
Sasaki:2016jop,Bird:2016dcv} 
(see \cite{macho,macho2} for earlier discussions).    

In the near future, we expect that a large number of binary BH systems 
will be detected 
as the sources of  gravitational wave events,
and we can discuss the origin of binary BH systems from a statistical perspective.
Among such statistical quantities,
the mass function and the spin distribution of the BHs should be useful~\cite{knn,Clesse:2016vqa}. The mass function of PBHs is studied in \cite{nj,jy,cks} 
by applying the critical phenomena near the threshold of 
the BH formation \cite{choptuik,ec,koike}. 
While the spin distribution of massive BH binaries formed from Population 
III stars has been recently studied in \cite{knn}, as far as we aware, 
little is known about the spin distribution of PBHs. 

Recently, Baumgarte and Gundlach \cite{bg,bg2}, 
motivated by analytical study by \cite{gundlach},  performed numerical simulations 
 of the collapse of a rotating radiation fluid and found the critical  
behavior of the angular momentum. 
It is of immediate interest to apply these findings to study  rotating PBHs.  
In this paper, we investigate the mass function and the spin distribution 
of the PBHs  
based on the recent study  of the critical phenomena in the formation of rotating BHs. 

This paper is organized as follows. In Sec. \ref{sec2}, first we formulate the density parameter of the PBHs, $\Omega_{\rm PBH}$, and 
then we derive  the mass function and the spin distribution of 
PBHs analytically for a simplified situation. We also study the evolution of the spin of a PBH after formation. Sec. \ref{sec3} is devoted to the conclusion. 
We use the units  $G=c=1$.

\section{Mass function and spin distribution}
\label{sec2}

%
Recently, the critical phenomena in the formation of rotating BHs from radiation fluids 
have been investigated \cite{bg,bg2}, 
and it has been found that  the angular momentum $J$ also obeys 
the scaling relation \cite{bg} and the mass and the angular momentum 
of the BHs  depend on the two parameter family of the initial data \cite{bg2}:
\beqa
M&=&C_M|\delta-\delta_c(q)|^{\gamma_M},\label{mass2}\\
J&=&C_J|\delta-\delta_c(q)|^{\gamma_J}q,\label{ang}\\
\delta_c(q)&=&\delta_{c0}+K q^2 \label{deltac}, 
\eeqa
where $\gamma_M\simeq 0.3558$ \cite{koike} and  
$\gamma_J=(5/2)\gamma_M\simeq 0.8895$.  The numerical value of 
$\gamma_J$ is  predicted analytically from the study of non-spherical perturbations of the critical (self-similar) solution \cite{gundlach} and 
is confirmed numerically \cite{bg}. 
Here $q$ denotes the parameter that characterizes 
the rotation of the initial data and $\delta_{c0}$ and $K$ are constants. 
Although the calculations are performed in  asymptotically flat spacetime, we 
expect that similar behavior exists in the collapse of radiation fluids in the expanding universe. 

 For demonstrative purposes, we limit ourselves to a Gaussian probability 
distribution function $P(\delta)$ for density fluctuations $\delta$ :
\beqa
P(\delta)=\frac{1}{\sqrt{2\pi} \sigma}\exp\left(-\frac{\delta^2}{2\sigma^2}\right),
\eeqa
 where $\sigma$ is the root-mean-square fluctuation amplitude and we assume  a flat distribution function $Q(q)$ for $q$.  
Then the volume fraction of 
the region collapsing into BHs at a given epoch is given by
\beqa
\beta(M_H)=\int^{\infty}_0 Q(q) dq \int^{\infty}_{\delta_c(q)}P(\delta)d\delta
\simeq \N\int^{\infty}_0 dq \frac{\sigma}{\sqrt{2\pi} \delta_c(q)}\exp 
\left(-\frac{\delta_c(q)^2}{2\sigma^2}\right),
\label{beta}
\eeqa
where $\N$ is a normalization constant and  we have performed 
the  integration with respect to $\delta$ by expanding $P(\delta)$ around $\delta = \delta_c(q)$.
The integral can be performed by expanding the integrand around $q=0$:
\beqa
\beta\simeq \frac{\N\sigma^2}{2\delta_{c0}\sqrt{2\delta_{c0}K}}\exp\left(-
\frac{\delta_{c0}^2}{2\sigma^2}
\right).
\eeqa
Here, we assume $\delta_{c0}/\sigma \gg 1$.
The density parameter of PBHs at a given epoch is then given by
\beqa
\Omega_{PBH}=\frac{\N}{M_H}\int^{\infty}_0 dq \int_{\delta_c(q)}
M(\delta)P(\delta)d\delta .
\label{om}
\eeqa

Let us introduce a dimensionless specific angular momentum (spin) parameter $a=J/M^2$ 
that corresponds to the dimensionless Kerr parameter.  From Eq. (\ref{mass2}) and 
Eq. (\ref{ang}), $a$ can be rewritten as  
\beqa
a=\frac{J}{M^2}=\frac{C_J}{C_M^2}|\delta-\delta_c(q)|^{\gamma_J-2\gamma_M}q=\frac{C_J}{C_M^2}\left(\frac{M}{C_M}\right)^{1/2}q,
\eeqa
where $\gamma_J=(5/2)\gamma_M$ is used in the last equality. 
$q$ and $\delta$ are now written in terms of $M$ and $a$: 
\beqa
q&=&\frac{C_M^2}{C_J}\left(\frac{M}{C_M}\right)^{-1/2}a,\\
\delta&=&\delta_c(q)+\left(\frac{M}{C_M}\right)^{1/\gamma_M}=
\delta_{c0}+K\left(\frac{C_M^2}{C_J}\right)^{2}\left(\frac{M}{C_M}\right)^{-1}a^2
+\left(\frac{M}{C_M}\right)^{1/\gamma_M}.
\eeqa
We can then make a change of variables from $(\delta,q)$ to $(M,a)$ in Eq. (\ref{om}) 
to calculate the mass-spin distribution function. 
Since the integration measure is transformed as
\beqa
dq d\delta=\frac{C_M}{\gamma_MC_J}\left(\frac{M}{C_M}\right)^{-3/2+1/\gamma_M}da dM,
\eeqa
we write $\Omega_{PBH}$ in terms  of $M$ and $a$ as
\beqa
\Omega_{PBH}&=&\frac{\N}{\sqrt{2\pi}\sigma M_H}
\frac{C_M^3}{\gamma_MC_J}\int_0 da \int_{-\infty} d\ln M \left(\frac{M}{C_M}\right)^{1/2+1/\gamma_M}\nonumber\\
&&\times\exp \left[-\frac{1}{2\sigma^2}\left(\delta_{c0}+K\left(\frac{C_M^2}{C_J}\right)^{2}\left(\frac{M}{C_M}\right)^{-1}a^2
+\left(\frac{M}{C_M}\right)^{1/\gamma_M}\right)^2\right].
\label{dom}
\eeqa

\subsection{Mass Function}

\begin{figure}
\includegraphics[height=3in]{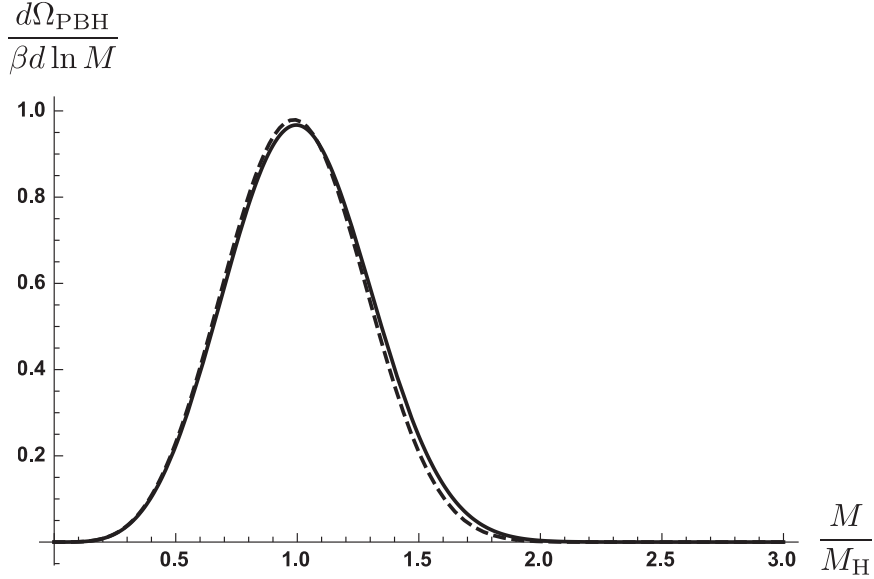}
\caption{\label{fig1} The mass function given by Eq. (\ref{massfunction}) 
is shown by  a solid curve. 
A dashed curve is obtained by performing the integration numerically.  
 }
\end{figure}

We can obtain the mass  function when we perform the $a$ integration in 
Eq. (\ref{dom}).  By expanding the integrand around $a=0$, the integral can 
be calculated to obtain the mass function
\beqa
\frac{d\Omega_{PBH}}{d\ln M}\simeq \frac{\N}{2\gamma_M\sqrt{2\delta_{c0}K}}
\frac{C_M}{M_H} \left(\frac{M}{C_M}\right)^{1+1/\gamma_M}
\exp\left[-\frac{1}{2\sigma^2}\left(\delta_{c0}^2+2\delta_{c0}
\left(\frac{M}{C_M}\right)^{1/\gamma_M}
\right)\right].
\eeqa
The mass function has a peak at $M_{max}=\left((1+\gamma_M)\sigma^2/\delta_{c0}\right)^{\gamma_M}C_M$ 
and drops steeply at $2M_{max}$. Since the horizon mass $M_H$ is 
the important mass scale in the problem and a typical black hole mass 
is around $M_H$, we equate $M_{max}$ with 
the horizon mass $M_H$~\cite{jy}.  Then we find
\beqa
\frac{d\Omega_{PBH}}{d\ln M}&=&\frac{\N(1+\gamma_M)}{2\gamma_M\sqrt{2\delta_{c0}K}}
\frac{\sigma^2}{\delta_{c0}} \left(\frac{M}{M_H}\right)^{1+1/\gamma_M}
\exp\left[-\frac{\delta_{c0}^2}{2\sigma^2}-(1+\gamma_M)
\left(\frac{M}{M_H}\right)^{1/\gamma_M}\right]\nonumber\\
&\simeq& \beta(M_H)\left(1+1/\gamma_M\right) 
\left(\frac{M}{M_H}\right)^{1+1/\gamma_M}\exp\left(-(1+\gamma_M)(M/M_H)^{1/\gamma_M}\right),
\label{massfunction}
\eeqa
where we have used Eq. (\ref{beta}). The mass function 
is the same as given in \cite{jy,nj}. 
The mass function is shown in Fig. \ref{fig1}. 

\subsection{Spin Distribution}

Next, we perform the $M$ integration  in Eq. (\ref{dom}) to obtain the spin distribution function:
\beqa
\frac{d\Omega_{PBH}}{da}&=&\frac{\N}{\sqrt{2\pi}\sigma M_H}
\frac{C_M^2}{\gamma_M C_J}\int_0 dM \left(\frac{M}{C_M}\right)^{-1/2+1/\gamma_M}
\nonumber\\
&&\times\exp\left[-\frac{1}{2\sigma^2}\left(\delta_{c0}+K\left(\frac{C_M^2}{C_J}\right)^{2}\left(\frac{M}{C_M}\right)^{-1}a^2
+\left(\frac{M}{C_M}\right)^{1/\gamma_M}\right)^2\right].
\eeqa
The integrand has a maximum at $M=M_*$, 
where $M_*$ for $a<1$ is given approximately by
\beqa
M_*\simeq \left(\left(1-\frac{\gamma_M}{2}\right)
\frac{\sigma^2}{\delta_{c0}}\right)^{\gamma_M}C_M .
\eeqa
Here, we assume $\delta_{c0} > (M_*/C_M)^{1/\gamma_M}$.
Then, by making use of the saddle point approximation
 around $M=M_*$ and further assuming $\delta/\sigma\gg1$, which 
may be the case for PBH formation \cite{nj,gl}, 
the integral can 
be calculated to obtain the spin distribution function:
\beqa
\frac{d\Omega_{PBH}}{da} &\simeq& {\N \over \sqrt{\delta_{c0}}}{C_M^2 \over C_J} {M_* \over M_H} \left({M_* \over C_M} \right)^{-1/2+1/(2\gamma_M)}
\nonumber\\
&&\times\exp\left[-\frac{1}{2\sigma^2}\left(\delta_{c0}+K\left(\frac{C_M^2}{C_J}\right)^{2}\left(\frac{M_*}{C_M}\right)^{-1}a^2
+\left(\frac{M_*}{C_M}\right)^{1/\gamma_M}\right)^2\right] \nonumber\\
&\simeq& \beta (M_H)\, {2 \delta_{c0} \sqrt{2K} \over \sigma^2}
 {C_M^2 \over C_J} {M_* \over M_H} \left({M_* \over C_M} \right)^{-1/2+1/(2\gamma_M)}
 \nonumber\\
&&\times\exp \left[-\frac{\delta_{c0}K}{\sigma^2}\left(\frac{C_M^2}{C_J}\right)^{2}\left(\frac{M_*}{C_M}\right)^{-1}a^2
-\frac{\delta_{c0}}{\sigma^2}\left(\frac{M_*}{C_M}\right)^{1/\gamma_M} \right]
\label{ang-distribution}
\eeqa
The distribution function is given approximately by a Gaussian function of $a$.

\begin{figure}
\includegraphics[height=3in]{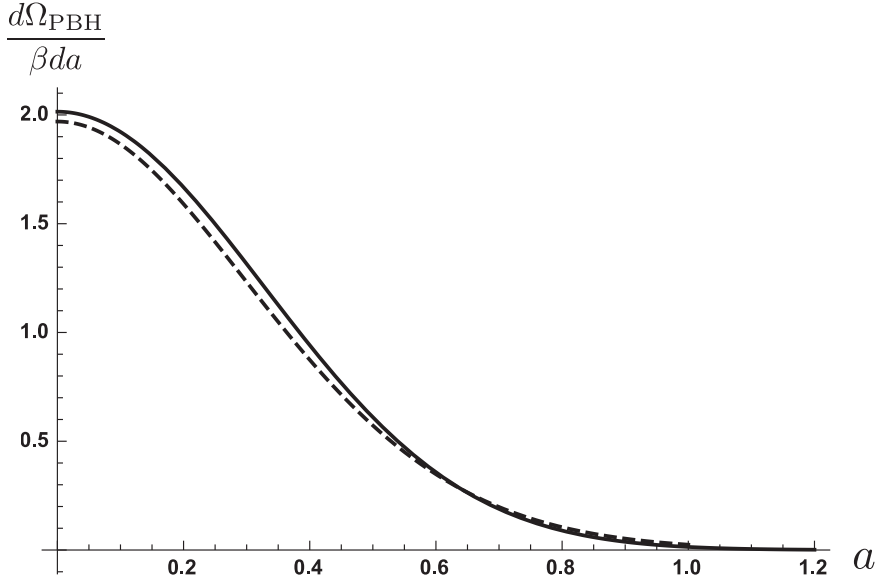}
\caption{\label{fig2} The spin distribution function given by Eq. (\ref{ang-distribution}) is shown by a solid curve.   
As in Fig.~\ref{fig1}, a dashed curve is obtained by performing the integration numerically.  
 }
\end{figure}

PBHs in cosmologically interesting numbers are formed 
in the early universe for $\sigma/\delta_{c0}\simeq 0.1-0.2$ \cite{gl,cksy}, which 
corresponds to $\beta(M_H)\simeq 5\times10^{-24}-4\times10^{-7}$.  
The value of  $\delta_{c0}$ (the threshold of the density perturbation in the 
comoving slice) is somewhat uncertain: 
Carr estimated $\delta_{c0}\simeq 1/3$ \cite{carr}. Recent numerical 
and analytical works suggest slightly larger value $\delta_{c0}\simeq 0.4\sim 
0.5$ \cite{shibata, musco1,harada}.  
For definiteness, we adopt  $\delta_{c0}=1/3$, $\gamma_M=0.3558$, and  
$\sigma/\delta_{c0}=0.15$, 
and  by  using the numerical values given  in \cite{bg2}, which corresponds to 
$C_M\simeq 5.118M_H,C_J\simeq 26.19 M_H^2$ and  $K\simeq 0.005685$ 
in our notation,\footnote{The derivation of these numerical values in our notation is given in Appendix.} 
we show the spin distribution function  in  Fig. \ref{fig2}. 
We find that PBHs formed in the early universe are mostly only slowly rotating, $a<0.4$. We note that the results do not change much for $\delta_{c0}=0.4$. 
Note also that the distribution function is not applicable near $a\simeq 1$ 
because the deviations from the scaling law are large \cite{bg2}.\footnote{Ref. \cite{bg} observed that Eq. (\ref{deltac}) is valid for $\Omega<0.3$, where 
$\Omega$ is the parameter controlling the angular momentum,  which corresponds to $a<0.7$ in our notation.}   
We have assumed a flat distribution for $q$ in deriving 
Eq. (\ref{ang-distribution}), for simplicity. 
For a  decreasing function $Q(q)$, 
$d\Omega/da$ would be much narrowed near $a=0$. 

\subsection{Spin Evolution}

In order to connect the spin of PBHs at  formation to that of the present time, 
we calculate the evolution of the spin of PBHs due to torque by the background 
 radiation fluid and mass accretion. 

A rotating BH sweeps the background radiation fluid and thus receives 
 momentum from the radiation \cite{hogan,macho2}. The force is estimated as
\beqa
F_{rad}&\simeq& ({\rm radiation~ momentum~ density})\times ({\rm cross~ section})\times ({\rm rotation~ velocity})\nonumber\\
&\simeq&\rho_{rad}\times M^2\times a,
\eeqa
where $\rho_{rad}$ is the energy density of the radiation. 
Then, the loss of angular momentum due to the torque of the force is estimated by
\beqa
\dot J\simeq - M F_{rad}\simeq -H^2M^3a,
\eeqa
where we have used the Friedmann equation $H^2\sim \rho_{rad}$. 
The mass of PBH grows due to the accretion of  background 
radiation on the PBH:
\beqa
\dot M\simeq \rho_{rad}M^2\sim H^2 M^2.
\label{massevolution}
\eeqa
The solution of Eq. (\ref{massevolution}) is given by \cite{ch}
\beqa
M\simeq \frac{t}{1+\frac{t}{t_i}(t_i/M_i-1)}, 
\eeqa
where $M_i$ is the mass at the initial time $t_i$.  Hence, for a PBH 
whose mass is lighter than the horizon mass,  $M_i< t_i$,  the mass of PBH 
grows little by accretion \cite{ch}. 

Since $M$ remains almost constant, the spin evolution is estimated as
\beqa
\dot a&=&\frac{d}{dt}\left(\frac{J}{M^2}\right)=\frac{\dot J}{M^2}-2\frac{J\dot M}{M^3}\nonumber\\
&\simeq& -H^2M_i a.
\eeqa
The equation can be integrated to obtain
\beqa
a\simeq a_i\exp\left(\alpha M_i/t_i\left(\frac{t_i}{t}-1\right)\right),
\eeqa
where $\alpha$ is a constant of $O(1)$. Hence,  the spin evolution is negligible 
for $M_i< t_i$,  $a\sim a_i$. \footnote{We note that 
spinning PBHs may suffer from 
super-radiant instability in the radiation dominated era because photons interacting with 
 a plasma  (free electrons) acquire an effective mass equal to 
the plasma frequency. 
Such a super-radiant instability due to cosmic plasma is effective for $M\siml 0.1 a M_{\odot}$\cite{pani}, and  a PBH satisfying this relation would lose its mass and angular momentum. }

\section{Conclusion}
\label{sec3}

Motivated by the recent study of the critical behavior of  angular 
momentum in the collapse of a rotating radiation fluid \cite{bg,bg2}, 
we calculate the distribution function of the spin of PBHs. We found that 
most PBHs are slowly rotating, $a<0.4$. 
This is basically because for larger $q$ the 
threshold density $\delta_c(q)$ becomes larger 
and hence the probability of  PBH formation is suppressed.
Note that the result should depend on the distribution function for $q$.
Here, we have assumed a flat distribution for $q$ for simplicity.
Naively, for PBHs the parameter $q$, which is related to  
the initial rotational mode,
can be calculated in the cosmological perturbation theory \cite{peebles},
and in the standard scenario the distribution for $q$ would be expected to have a peak around $q = 0$. 
Hence, we expect that the spin distribution of PBHs is much narrowed 
near $a=0$ in more realistic situation.  
However, it is important and should be interesting to investigate the distribution function of the initial rotational mode,
and we leave it as a future study. 
We also estimate the spin evolution after the formation
and find that it is expected to be negligible.

The evolution of  Population III star binaries, which could be 
the sources of  $30\,{\rm M}_\odot$ BH binaries, has been recently 
studied by \cite{knn},  and 
it is found that the spins of a large fraction of the resulting BHs are 
high: $a\sim 1$. Therefore, we expect that 
the spins of binary BHs can be the probe of the origin 
of binary BHs.

\section*{Acknowledgments}

This work is supported by MEXT KAKENHI Grant Number 15H05894 (TC), 15H05888 (SY) and  16H01103 (SY) and by JSPS KAKENHI Grant Number
15K17659 (SY) and in part by Nihon University (TC). We thank Teruaki Suyama and Jun'ichi Yokoyama for useful comments and Paolo Pani for useful information.

\section*{Appendix: Conversion of Units}
\label{sec:app}

In order to plot $d \,\Omega_{\rm PBH} / da$ as a function of $a$,
here we show the values of input parameters ($C_M, C_J, K, \delta_{c0}, \sigma, \gamma_M,$ and $\gamma_J$) which we use in the integration with respect to $M$.
Except for the values of $\gamma_J$, $C_J$ and $K$, we can use the values of input parameters
in the case  $q = 0$. Following Refs.~\cite{koike, nj, jy, bg, bg2},
we use $\delta_{c0} = 1/3$, $\sigma / \delta_{c0} = 0.15$ and $\gamma_M = 0.3558 $.
In Ref. \cite{jy}, it was found that
in the case  $q = 0$, the mass function has a peak at 
$M_{\rm max} := C_M (\delta_{c0} / \sigma^2)^{-\gamma_M} (1 + \gamma_M)^{\gamma_M}$
and $M_{\rm max}$ can be identified with $M_H$.
Following this paper, we can estimate
\begin{eqnarray}
C_M = (\delta_{c0} / \sigma^2)^{\gamma_M} (1 + \gamma_M)^{-\gamma_M} M_H \simeq  5.117 \, M_H.
\label{eq:JY_M}
\end{eqnarray}

Next, let us consider the parameters related to angular momentum, following Ref. \cite{bg2}.
Eq. (20) in \cite{bg2}, given by
\begin{eqnarray}
M = C_0^{\gamma_M} (\eta - \eta_{\ast 0})^{\gamma_M},
\label{eq:GB_M}
\end{eqnarray}
is for the $\Omega = 0$ sequence, where $\Omega$ is a control parameter related to 
  angular momentum
and hence it corresponds to $q$ in our notation.
As shown in Eq. (20), $\eta_{\ast 0} = 1.0183772$ and $C_0 = 0.28$.
Comparing our notation with the above formula, we can find that
we have a relation between $\delta$ and $\eta$
\begin{eqnarray}
\eta = A \, \delta,~{\rm with}~A := \eta_{\ast 0} / \delta_{c0} \simeq 3.05513.
\label{eq:GB_our}
\end{eqnarray}
Substituting this relation into  (\ref{eq:GB_M}), we have
\begin{eqnarray}
M = \left( C_0 A \right)^{\gamma_M} \left( \delta - \delta_{c0} \right)^{\gamma_M},
\end{eqnarray}
and then $C_M$ is given by $\left( C_0 A \right)^{\gamma_M} \simeq 0.946$.
However, we do not know the overall normalization scale in Ref.~\cite{bg2},
and here we denote it by $M_{\rm GB}$, i.e.,  $C_M =  \left( C_0 A \right)^{\gamma_M} \simeq 0.946 \, M_{\rm GB}$.
By comparing this expression with Eq. (\ref{eq:JY_M}), $M_{\rm GB} \simeq 5.41 \, M_H$. 

Based on the above expression,
let us evaluate $C_J$ and $ K$ in our notation. 
Eq. (21b) in \cite{bg2} is given by
\begin{eqnarray}
J = \left( \bar{\eta} - \bar{\eta}_\ast \right)^{\gamma_J} \bar{\Omega_\ast},
\end{eqnarray}
where $\bar{\eta} = C_0 ( \eta - \eta_{\ast 0})$ (from Eq. (20) in \cite{bg2}) and $\bar{\eta}_\ast = K_{\rm GB} \bar{\Omega}_\ast^2$
has been used. Let us rewrite the above equation in terms of $\delta$ in our notation by using Eq. (\ref{eq:GB_our}).
\begin{eqnarray}
J = \left( \bar{\eta} - \bar{\eta}_\ast \right)^{\gamma_J} \bar{\Omega_\ast} &=& C_0^{\gamma_J}
\left( \eta - \eta_{\ast 0} - {K_{\rm GB} \over C_0 } \bar{\Omega}_\ast^2 \right)^{\gamma_J} \bar{\Omega}_\ast \cr\cr
&=& \left( C_0 \, A \right)^{\gamma_J} \left( \delta - \delta_{c0} - {K_{\rm GB} \over C_0 A } \bar{\Omega}_\ast^2 \right)^{\gamma_J} \bar{\Omega}_\ast .
\end{eqnarray}
As shown  above, $(C_0\,A)^{\gamma_M} = 0.946 \, M_{\rm GB}$, and hence
$\left( C_0 \, A \right)^{\gamma_J} = \left( 0.946 \, M_{\rm GB} \right)^{\gamma_J / \gamma_M}$.
By using $\gamma_J / \gamma_M = 5/2$, we have $\left( C_0 \, A \right)^{\gamma_J} \simeq 0.8704 \, M_{\rm GB}^{5/2}$.
However, the dimension of $J$ is $[{\rm mass}^2]$, and hence $\bar{\Omega}_\ast$ should have $[{\rm mass}^{-1/2}]$.
Introducing a dimension less parameter $q$ as
\begin{eqnarray}
q := \bar{\Omega}_\ast \times  \left( C_0 \, A \right)^{\gamma_M / 2},
\end{eqnarray}
the above equation for $J$ can be written as
\begin{eqnarray}
J &=& \left( C_0 \, A \right)^{\gamma_J} \left( \delta - \delta_{c0} - {K_{\rm GB} \over C_0 A } \bar{\Omega}_\ast^2 \right)^{\gamma_J} \bar{\Omega}_\ast
\cr\cr
&=& \left( C_0 \, A \right)^{ \gamma_M \left({\gamma_J \over \gamma_M} - {1 \over 2}\right) }
\left( \delta - \delta_c^0 - {K_{\rm GB} \over \left( C_0 \, A \right)^{\gamma_M + 1} } q^2 \right)^{\gamma_J} q \cr\cr
&=& C_J \left( \delta - \delta_{c0} - K q^2 \right)^{\gamma_J} q,
\end{eqnarray}
with $K \simeq 0.005685 $ and $C_J = \left( C_0 \, A \right)^{ 2 \gamma_M} \simeq 0.8949 \, M_{\rm GB}^2 \simeq 26.19 \, M_{H}^2$.


\end{document}